\documentstyle [twocolumn,prl,aps,floats,epsfig] {revtex}
\begin{document}
\draft
\begin{title}
{Collective and independent-particle motion in two-electron artificial atoms}
\end{title} 
\author{Constantine Yannouleas and Uzi Landman} 
\address{
School of Physics, Georgia Institute of Technology,
Atlanta, Georgia 30332-0430 }
\date{Phys. Rev. Lett. {\bf 85}, 1726 (2000); Received 14 March 2000}
\maketitle
\begin{abstract}
Investigations of the exactly solvable excitation spectra of two-electron
quantum dots with a parabolic confinement, for different values of the
parameter $R_W$ expressing the relative magnitudes of the interelectron
repulsion and the zero-point kinetic energy, reveal 
for large $R_W$ a ro-vibrational spectrum associated
with a linear trimeric rigid molecule composed of the two 
electrons and the infinitely heavy confining dot. This spectrum transforms to 
that of a ``floppy'' molecule for smaller values of $R_W$. The 
conditional probability distribution calculated for the exact two-electron 
wave functions allows for the identification of the ro-vibrational excitations
as rotations and stretching/bending vibrations, and provides direct evidence 
pertaining to the formation of such molecules.
\end{abstract}
\narrowtext
The behavior of three-body systems has been a continuing subject of interest
and a source of discoveries in various branches of physics, both in the
classical and quantum regimes, with the moon-earth-sun system \cite{gutz} and 
helium-like atoms \cite{kh,berr,rost,wrt,ostr,lin} (in the ground and excited 
states) being perhaps the best known examples. Furthermore, insights gained 
through such investigations often provide the foundations for understanding 
the properties of systems with a larger number of interacting particles.

Recently, analysis of the measured conductance \cite{taru} and differential 
capacitance \cite{asho} spectra of two-dimensional (2D) quantum dots
(QD's,) created via voltage gates at semiconductor heterointerfaces, led to 
their naming (by analogy) as ``artificial atoms''. In particular, 
this analogy refers to identification of regularities in the measurements  
which have been interpreted \cite{taru} along the lines of the electronic 
shell model (SM) of natural atoms, which is founded on the physical picture of
electrons moving in a {\it spherical central\/} field including the 
averaged contribution from electron-electron interactions.

Motivated by the central role that spectroscopy played in the development 
of our undestanding of atomic structure, we investigate in this paper the 
exactly solvable excitation spectrum of a two-electron (2e) parabolic QD as a 
prototypical three-body problem comprised of the
two electrons ($X$'s) and the (infinitely heavy) confining quantum dot ($Y$). 
Through probing of the structure of the exact wave functions  with the use of
the conditional probability distribution (CPD) \cite{berr}, in conjunction 
with identification of regularities of the excitation spectrum,
we show that such a spectrum is characteristic of collective 
dynamics resulting from formation of a linear trimeric molecule $XYX$
\cite{note1}. In particular, we find that the excitation spectrum of the 2e QD
exhibits for a weak parabolic confinement (i.e., small harmonic
frequency $\omega_0$) a well-developed, separable ro-vibrational pattern
which is akin to the characteristic spectrum of natural ``rigid'' triatomic
molecules (i.e., molecules with stretching and bending vibrational frequencies
higher than the rotational one). For stronger confinements (i.e., large 
$\omega_0$), the spectrum transforms to one characteristic of a ``floppy'' 
triatomic molecule, converging finally to the independent-particle picture 
associated with the circular central mean field of the QD. 

The Schr\"{o}dinger equation for a 2e QD with a parabolic confinement 
of frequency $\omega_0$, with the 2D Hamiltonian given by 
$H= \sum_{i=1,2} {\bf p}^2_i/2m^* + e^2/\kappa |{\bf r}_1 - {\bf r}_2|
+0.5 m^* \omega_0^2 \sum_{i=1,2} {\bf r}_i^2$,
where $\kappa$ and $m^*$ are, respectively, the dielectric constant and 
electron effective mass, is separable in 
the center-of-mass (CM) and relative-motion (rm) 
coordinates \cite{garc}. Consequently, the energy eigenvalues may be
written as $E_{NM,nm}=E^{\text{CM}}_{NM} + \varepsilon^{\text{rm}}(n,|m|)$,
where $E^{\text{CM}}_{NM}=\hbar \omega_0 (2N+|M|+1)$ with the $N$ and $M$
quantum numbers corresponding to the number of radial nodes in the CM
wave function and $M$ is the CM azimuthal quantum number, and 
$\varepsilon^{\text{rm}}(n,|m|)$ are the eigenvalues of the one-dimensional
Schr\"{o}dinger equation \cite{garc},
\[
\frac{\partial^2 \Omega}{\partial u^2} + 
\{ \frac{-m^2+1/4}{u^2}-u^2-\frac{R_W \sqrt{2}}{u}+
\frac{\varepsilon}{\hbar \omega_0 /2} \} \Omega =0~,
\]
where $\Omega(u)/\sqrt{u}$ is the radial part of the rm wave function
$\Omega(u)e^{im\theta}/\sqrt{u}$ with $n$ being the number of radial nodes;
$u=|{\bf u}_1-{\bf u}_2|$ with ${\bf u}_i={\bf r}_i/l_0 \sqrt{2}$ $(i=1,2)$
being the electrons' coordinates in dimensionless units and 
$l_0=(\hbar/m^* \omega_0)^{1/2}$, that is the spatial extent of the 
lowest-state wave function of a single electron. The so-called Wigner 
parameter $R_W=(e^2/\kappa l_0)/\hbar \omega_0$ multiplying the Coulomb 
repulsion term expresses the relative strength of the Coulomb repulsion 
between two electrons separated by $l_0$ and twice the zero-point kinetic 
energy of an electron moving in a harmonic confinement.

Denoting the exact spatial wave function of the 2e QD by 
$\Phi_{NM,nm}({\bf u}_1,{\bf u}_2)$ (which is the product of the CM and rm 
wave functions), and the spatial two-electron density by
$W_{NM,nm}({\bf u}_1,{\bf u}_2)=|\Phi_{NM,nm}({\bf u}_1,{\bf u}_2)|^2$,
we define the usual pair-correlation function (PCF) as
\[
G(v)=2\pi \int \int \delta({\bf u}_1-{\bf u}_2-{\bf v}) 
W({\bf u}_1,{\bf u}_2) d{\bf u}_1 d{\bf u}_2~,
\]
and the conditional probability distribution (CPD) for finding one 
electron at ${\bf v}$ given that the other is at ${\bf v}_0$ as,
\[
{\cal P}({\bf v}|{\bf u}_2={\bf v}_0)=
\frac{W({\bf v},{\bf u}_2={\bf v}_0)}
{\int d{\bf u}_1 W({\bf u}_1,{\bf u}_2={\bf v}_0)}~,
\]

\newpage
\widetext
\begin{figure}[t]
\centering\epsfig{file=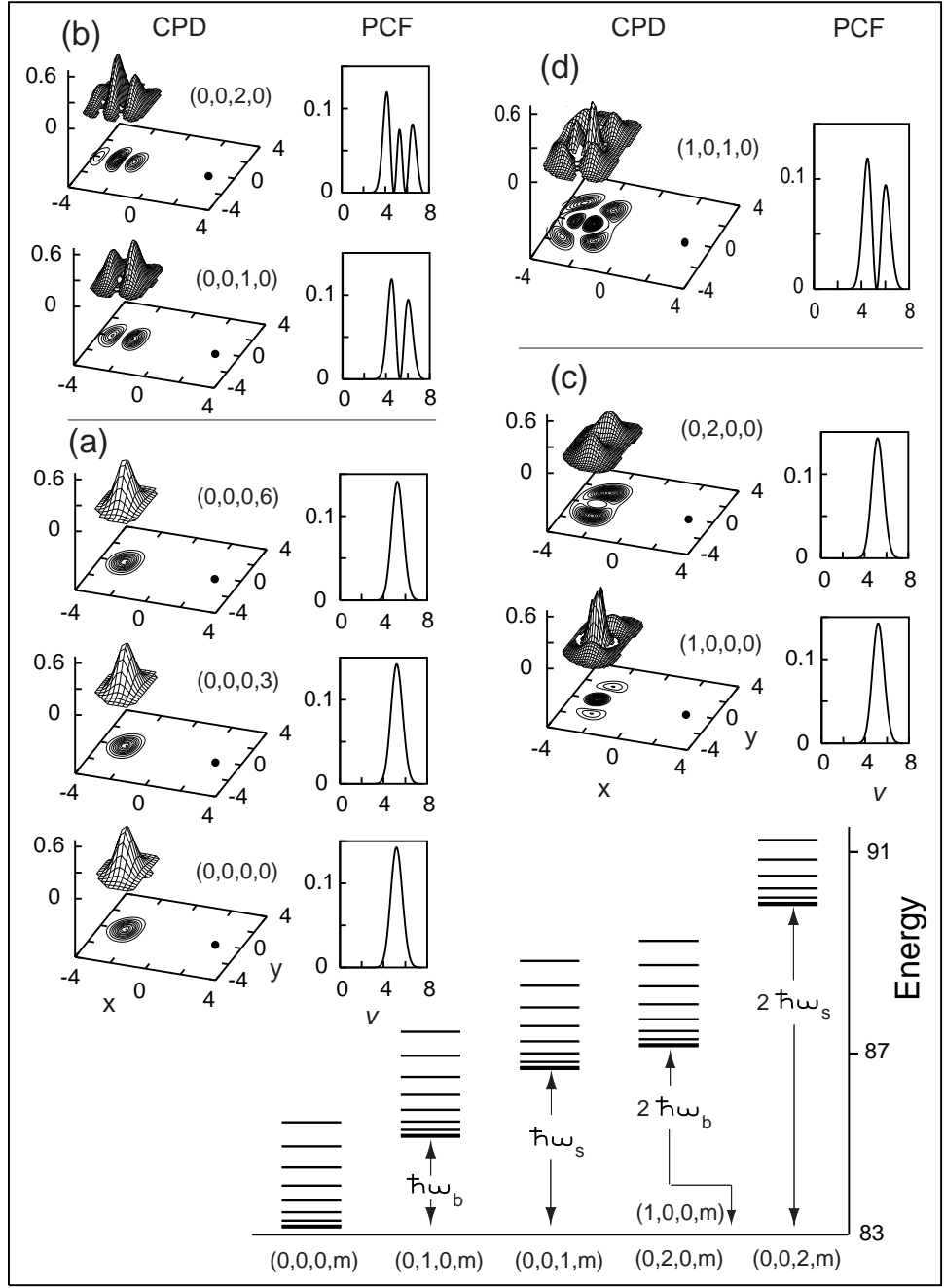,%
width=14.cm,clip=,angle=0}\\
~~~~~~~~~~~\\
\caption{
Spectra (bottom) and corresponding conditional probability distributions
(CPD) and pair-correlation functions (PCF) shown respectively on the left and 
right of each of the subplots [labeled (a)-(d)], for a 2eQD with $R_W=200$.
For each excitation band, the quantum numbers $(N_0,M_0,n_0,m)$ are given
at the bottom with $m=0,1,2,...$ (the levels for $m=0$ and $m=1$ are not
resolved on the scale of the figure and appear as a thick line);
only a few of the low lying rotational and vibrational states are shown,
with the collective rovibrational behavior extending to higher excitations.
The CPD's and PCF's are labeled with the
quantum numbers of the corresponding levels. For the spectral rules governing
the spectrum and for the definition and interpretation of the CPD's and PCF's,
see the text. The solid dot in each of the CPD subplots denotes the point
${\bf v}_0=(d_0,0)$, where $d_0=2.6$  is half of the electron separation 
found in the PCF of the ground state (0,0,0,0). All distances $x,y,v$ and
$d_0$ are in units of $l_0 \sqrt{2}$, and energies are in units of 
$\hbar \omega_0/2$.
}
\end{figure}
\narrowtext
\newpage
~~~~~~~\\
\newpage
~~~~~~~\\
\newpage
where the $M,N,n,m$ indices of $W$ (and therefore of $G$ and ${\cal P}$) have
been suppressed. Note that the exact electron densities are circularly
symmetric.

With the above, we solved for the 2e QD energy spectra and wave functions
for values of $R_W=200$, 20 and 3. We discuss first the $R_W=200$ case
whose spectrum and selected PCF's and CPD's are displayed in Fig.\ 1.
As can be seen immediately, for such a large value of $R_W$, the spectrum
of the 2e QD (bottom part of Fig.\ 1) exhibits the following three 
well-developed regularities: 
(I) for every band $(N_0,M_0,n_0,m)$, with $m=0,1,2,... $, while 
$N_0,M_0$ and $n_0$ are kept constant (in the following the subscript ``zero''
denotes a number that is held constant in a particular sequence), 
the energy spacing between two adjacent levels $m$ and $m+1$ {\it increases 
linearly \/} in proportion to $2m+1$; the bands $(N_0, \pm M_0, n_0, \pm m)$ 
are degenerate. Note that the levels are spin singlet or triplet for $m$ even 
or odd, respectively, (II) the bands $(0,M_0,0,m)$ and $(N_0,0,0,m)$ 
correspond to excitations of the center-of-mass motion with $M_0$ and $2N_0$ 
vibrational quanta (phonons) of energy $\hbar \omega_0$, respectively, and
(III) the bottom levels of the bands $(0,0,n_0,m)$ form a one-dimensional
harmonic-oscillator spectrum $(n_0+1/2) \hbar \omega_s$.

The above three ``spectral rules'' specify a well-developed and separable 
ro-vibrational spectrum exhibiting collective rotations, as well as 
stretching and bending vibrations \cite{note4}. Indeed, neglecting an overall 
constant term, the above rules can be summarized as,
\[
E_{NM,nm}=C m^2~ + (n+1/2) \hbar \omega_s + (2N+|M|+1) \hbar \omega_b~,
\]
where the rotational constant $C \approx 0.037$, the phonon for the
stretching vibration has an energy $\hbar \omega_s \approx 3.50$, and the 
phonon for the bending vibration coincides with that of the CM motion, i.e., 
$\hbar \omega_b = \hbar \omega_0=2$ \cite{note4,note5} (all energies are given
in dimensionless units of $\hbar \omega_0/2$). Note that the rotational energy
is proportional to $m^2$, as is appropriate for 2D rotations, unlike the case 
of natural triatomic molecules where the rotational energy has a term
proportional to $l(l+1)$, $l$ being the quantum number associated with
the 3D angular momentum. Observe also that the bending vibration can
carry by itself an angular momentum $\hbar M$ and thus the rotational angular 
momentum $\hbar m$ does not necessarily coincide with the total angular 
momentum $\hbar (M+m)$.
 
Further insight into the collective character of the spectrum displayed 
in Fig.\ 1 can be gained by examining the CPD's and PCF's associated with
selected states of the rotational bands $(N_0,M_0,n_0,m)$ (the CPD's are 
displayed to the left of the PCF's; notice that the PCF's are always 
circularly symmetric). The band $(0,0,0,m)$, being {\it purely\/} rotational 
with zero phonon excitations, can be designated as the ``yrast'' band, in 
analogy with the customary terminology from the spectroscopy of rotating 
nuclei \cite{bm}.

In Fig.\ 1(a), we display the CPD's and PCF's for three specific states of 
the yrast band, i.e., the (0,0,0,0), the (0,0,0,3), and the (0,0,0,6). The 
corresponding PCF's are all alike and centered around $2d_0=5.2$, which
implies that the two electrons keep apart from each other at a distance
$2d_0$. Due to the circular symmetry of the PCF's, however, one can only
conclude that the two electrons are moving on a thin circular shell of radius
$d_0$. To reveal the formation of an electron molecule, one needs to
consider further the corresponding CPD's [plotted in the left column with
${\bf v_0}=(d_0,0)$; the point ${\bf v}_0$ is denoted by a solid dot]. In 
fact, the CPD's demonstrate that the two electrons reside at all instances at 
diametrically opposite points, thus forming a linear molecule $XYX$ with two 
equal bonds ($X-Y$ and $Y-X$) of length $d_0$. In 
addition, one can see that all three CPD's are practically identical, in spite
of the fact that the angular momentum changes from $m=0$ (lower subplot) to 
$m=6$ (upper subplot). This behavior, namely the constancy of the bond lengths
irrespective of the rotational energy, properly characterizes the electron
molecule as a rigid rotor. 

Turning our attention away from the yrast band, we focus next on the bands
$(0,0,1,m)$ and $(0,0,2,m)$, which are rotational bands built upon one- and
two-phonon excitations of the stretching vibrational mode. We have verified
that the PCF's and the CPD's corresponding to these bands share with the
yrast band the property that they do not change (at least for the levels 
displayed in Fig.\ 1) as a function of $m$. Thus it is sufficient to study
the bottom states, i.e., those with $m=0$, (0,0,1,0) and (0,0,2,0), whose 
corresponding PCF's and
CPD's are displayed in the lower and upper subplots of Fig.\ 1(b),
respectively. The PCF's demonstrate the presence of internal
excitations with one and two nodes in the relative motion, but they yield no
further information regarding the electron molecule. The CPD's, however,
plotted here for ${\bf v}_0=(d_0,0)$ [the point ${\bf v}_0$ is kept the
same for all subplots in Fig.\ 1] immediately reveal the presence of 
excitations (specified by the number of their nodes, i.e., here one or two) 
associated with the vibrational mode of the $XYX$ molecule {\it along\/} the 
interelecton axis (namely, the stretching vibration).

By examining the corresponding CPD's, one can further demonstrate that 
the two degenerate rotational bands $(0,2,0,m)$ and $(1,0,0,m)$ are built
upon the lowest two-phonon excitations of the bending vibrational mode of the
linear molecule $XYX$. Again, we have verified that it is sufficient to 
consider the two states at the bottom of the bands, namely the $(1,0,0,0)$ 
[see lower subplot of Fig.\ 1(c)] and the $(0,2,0,0)$ [see upper subplot of 
Fig.\ 1(c)]. It can be seen that both CPD's describe vibrational 
excitations of the $XYX$ molecule which are {\it perpendicular\/} to the 
interelectron axis (namely, bending vibrations), with the one associated 
with the $(1,0,0,0)$ level having one node 
and the one associated with the $(0,2,0,0)$ 
having no nodes (this is in agreement with the fact that the normal mode 
associated with the bending vibrations is related to the 2D
harmonic-oscillator describing the CM motion). We note that the corresponding
PCF's [see right column in Fig.\ 1(c)] fail to describe (in fact they are
completely unrelated to) the bending vibrations; indeed they are identical to
the ones associated with the yrast band [Fig.\ 1(a)] which is devoid of
any vibrational excitations.

The CPD and PCF of the bottom level (i.e., with $m=0$) of the rotational band 
$(1,0,1,m)$, which is built upon more complicated phonon excitations of mixed 
bending and stretching character (not shown in Fig.\ 1), are displayed in 
Fig.\ 1(d). It is easily seen that the CPD represents a vibrational motion of 
the electron molecule both along the interelectron axis (one excited 
stretching-mode phonon) and perpendicularly to this axis (two excited 
bending-mode phonons). In fact, the CPD in Fig.\ 1(d) can be viewed as a 
composite made out of two CPD's shown previously, one in the lower subplot of 
Fig.\ 1(b) and the other in the lower subplot of Fig.\ 1(c). Returning to Fig.\
1(d), one can see again that, in contrast to the CPD which enables detailed
probing of the excitation spectrum, the information which may be extracted
from the corresponding PCF is rather limited.

\begin{figure}[t]
\centering\epsfig{file=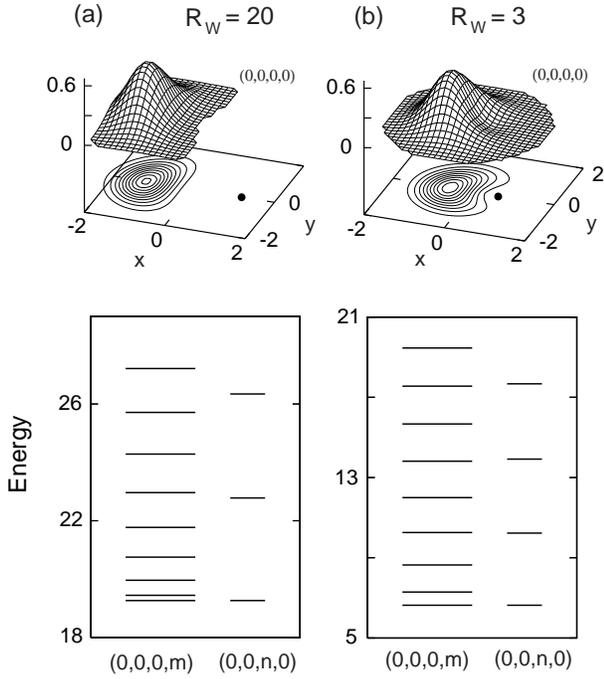,%
width=8.cm,clip=,angle=0}
\caption{
Spectra (bottom) and CPD's (top) of the corresponding ground-state levels
for 2eQD's with (a) $R_W=20$, and (b) $R_W=3$. At each of the bottom panels,
the spectrum of the yrast band [i.e., $(0,0,0,m)$; $m=0,1,2,...$] is shown
on the left, and the lowest levels of the bands $(0,0,n_0,m)$ for $n_0=0,1,2,
...$, forming a stretching vibrational spectrum (i.e., with constant spacing)
are displayed on the right. The solid dot in each of the CPD subplots denotes 
the point ${\bf v}_0=(d_0,0)$, where $d_0$ is half of the electron separation 
found in the corresponding ground-state PCF's (not shown here). Energies in 
units of $\hbar \omega_0/2$ and distances in units of $l_0 \sqrt{2}$; e.g.,
for GaAs material parameters ($m^*=0.067m_e$, $\kappa=12.9$) and $R_W=3$,
one has $\hbar \omega_0=1.2$ meV and $l_0 \sqrt{2}=43.54$ nm. 
}
\end{figure}
The rigidity of the electron molecule, which is so well established for
$R_W=200$, will naturally weaken as the parameter $R_W$ decreases and the
XYX molecule will start exhibiting an increasing degree of ``floppiness''.
Such floppiness can be best observed in the yrast band, which, beginning with
the higher levels, will gradually deviate from the spectral rule (I) discussed
above, and eventually it will become unrecognizable as a rotational band. This
is illustrated in the lower subplot of Fig.\ 2(a) which displays the yrast band
for $R_W=20$. Specifically, one can see that only the lowest four levels
honor approximately rule (I), the higher ones tending to develop a 
constant energy spacing between adjacent levels [this spacing converges
slowly to the energy spacing $\hbar \omega_0$ (i.e., to the value 2 in 
dimensionless units) of the parabolic confinement].
In the case $R_W=3$, one can hardly identify any rotational sequence in the 
levels of the yrast band [plotted at the bottom subplot of Fig.\ 2(b)]. Indeed,
although the energy spacing between the second and the third levels is larger 
than that between the first and the second levels (but with a ratio 
substantially different than 3/1), the spacing between higher levels approaches
quickly the value 2 of the external confinement. 

However, in spite of the floppiness exhibited by the excitation spectra in
Fig.\ 2, the (singlet) ground-state of the 2e QD for both $R_W=20$ and $R_W=3$
drastically deviates from the 1$s^2$ closed-shell orbital configuration 
expected from the independent-particle picture. Rather, as demonstrated by the
corresponding CPD's [top subplots in Fig.\ 2], in both these cases of smaller
$R_W$'s, the ground state is still associated with formation of rather
well-developed XYX electron molecules, but with progressively smaller
bond lengths. Finally, we remark that the stretching vibrations are more
robust and tend to better preserve a constant spacing between the bottom
levels of the bands $(0,0,n_0,m)$ [these levels were grouped in a vibrational
band $(0,0,n,0)$ and are plotted on the right-hand-side of the lower subplots
in Fig.\ 2].

The remarkable emergence of ro-vibrational excitations for parabolically
confined 2eQD's, under magnetic-field-free conditions, provides direct 
evidence for the formation of electron molecules in QD's, with their rigidity 
controlled by the parameter $R_W$. Such electron molecules and associated 
collective excitation spectra are general properties \cite{cy,cy1,maks} of 
QD's (with greater spectral complexity in many-electron QD's), whose 
observations (and manipulations through controlled pinning of the collective 
rotations[15(b)]) form outstanding experimental challenges.

This research is supported by the US D.O.E. (Grant No. FG05-86ER-45234).


\begin{references}
\bibitem{note1}
Molecular behavior of a nonrigid linear structure undergoing collective
rotation and vibration has been studied in the context of excited states
of two-electron atoms (Refs.\ \cite{kh,berr,rost,wrt,ostr,lin}).
\bibitem{cy1}
Formation of electron molecules in QD's has been 
discussed in Ref.\ \cite{cy} in the context of symmetry-breaking states of the
self-consistent field resulting from spin-and-space unrestricted Hartree-Fock 
calculations.
\end{references}
\end{document}